\documentclass[12pt]{iopart}

\usepackage{graphics}

\newcommand{\bea}{\begin{eqnarray}}
\newcommand{\eea}{\end{eqnarray}}
\newcommand{\be}{\begin{equation}}
\newcommand{\ee}{\end{equation}}

\def\be{\begin{eqnarray}}
\def\ee{\end{eqnarray}}
\def\bd{\begin{displaymath}}
\def\ed{\end{displaymath}}

\begin{document}

\title{Valence particles and the correction to relativistic mean field binding 
energy} 

\author{G. Gangopadhyay}
\address{Department of Physics, University of Calcutta\\
92, Acharya Prafulla Chandra Road, Kolkata-700 009, India\\
email: ggphy@caluniv.ac.in}
\date{}

\begin{abstract}

The differences between the experimental and the theoretically
calculated binding energies in Relativistic Mean Field (RMF) approach have been 
calculated for a large number of odd-$Z$ nuclei from $A=47$ to 229.   
Neutron-proton (n-p) interaction is expected to be the major contributor to 
this difference. This difference, excluding certain mass regions and taking 
other effects as well as the odd-even mass difference into account, may be 
linearly parametrized by the Casten factor, a recognized measure of the n-p 
interaction in the nucleus. The results follow the same pattern as in the case 
of even-$Z$ nuclei observed earlier. 
\end{abstract}

 
\maketitle

\section{Introduction}

The effective numbers of valence particles (or holes)
are often found to be useful in the parametrization of
various nuclear quantities\cite{Casten1}. The product of numbers of valence 
protons ($N_p$) and neutrons ($N_n$),  or similar functions of $N_p$ and $N_n$,
represent the  integrated n-p interaction strength  and hence have been found 
to bear smooth relationships with certain observables such as deformation and 
B(E2) values\cite{Casten3,Foy,Zhao}, properties of excited states
\cite{Casten2,yoon}, rotational moments of inertia and  ground band energy 
systematics\cite{Saha,Saha1},  spectroscopic factors \cite{plb2,jpg} etc. 
In Ref. \cite{plb2}, spectroscopic factors and the contribution of the n-p 
interaction to binding energy in 
actinides were seen to follow a certain pattern. In actinides, the only 
appropriate major doubly closed shell nucleus is $^{208}$Pb and it was necessary
to employ subshell closures. In another communication \cite{PLB0}, we obtained
a more robust systematic behaviour in the latter quantity in even-proton nuclei, 
valid in a large mass region and dependent 
only on the known major shell closures. In the present work, we extend our study to 
odd-proton nuclei.

As was pointed out in \cite{PLB0}, the correlations beyond mean field results 
are due principally to residual two body interaction. In mean field 
calculations, while the residual interaction between similar nucleons is  
taken care of by the introduction of $T=1$ pairing, the residual n-p interaction
is often ignored. 
The difference between the experimental and the calculated binding energies, 
suitably corrected for effects not related to the n-p interaction including odd-even
mass difference, may vary smoothly as a function  of the integrated strength 
of the n-p interaction. This difference, thus, may be expected to scale as 
the Casten factor $P=N_pN_n/(N_p+N_n)$, \cite{P} a widely used measure of the n-p 
interaction strength. 

\section{Calculation and Results}

Following the method in \cite{PLB0}, the effect of the n-p interaction in
the difference between the experimental and the theoretical binding energies 
has been extracted by assuming that, in a nucleus with magic neutron number,
this difference is due to the combined effects other than the n-p interaction.
As $N_n$ is zero in these particular nuclei, the effect of the n-p interaction
is expected to be small. The difference between theory and experiment 
in the change in the binding energy from the isotope with $N_n=0$ for a 
particular Z is taken as a measure of the contribution of the n-p interaction 
and expressed as $\Delta_{\nu\pi}$. Thus we write
\be \Delta_{\nu\pi}(Z,N)=A(B_{th}(Z,N)-B_{ex}(Z,N)+B_{corr}(Z))\ee
where, $B_{th}$ and $B_{ex}$ are respectively the theoretically
calculated and experimentally measured binding energies per nucleon and, 
$A=Z+N$, the mass number. We have defined  
$B_{corr}(Z)=B_{ex}(Z,N_0)-B_{th}(Z,N_0)$, 
$N_0$ being a magic number. The quantity $\Delta_{\nu\pi}(Z,N)$ is defined to
be zero in magic $N$ nuclei. The 
experimental binding energy values are from Ref. \cite{mass}.
This difference also incorporates the odd-even mass difference in odd-$N$ 
nuclei.

The choice of appropriate Lagrangian density is not unique as there are 
different variations of the Lagrangian density as well different 
parametrizations for them in RMF.  In Ref. \cite{PLB0}, we employed two such 
densities FSU Gold\cite{prl} and NL3\cite{NL3} and obtained nearly identical 
behaviour. Accordingly, most of the results presented here use only the former 
one. This density involves self-coupling of the vector-isoscalar meson as well 
as coupling between the vector-isoscalar meson and the vector-isovector meson.
The FSU Gold Lagrangian density seems very appropriate
for a large mass region {\em viz}. medium mass to superheavy nuclei. 
We have solved the equations in co-ordinate space. The strength of the zero 
range pairing force is taken as 300 MeV-fm for both
protons and neutrons. 

\begin{figure}
\center
\resizebox{7.4cm}{!}{ \includegraphics{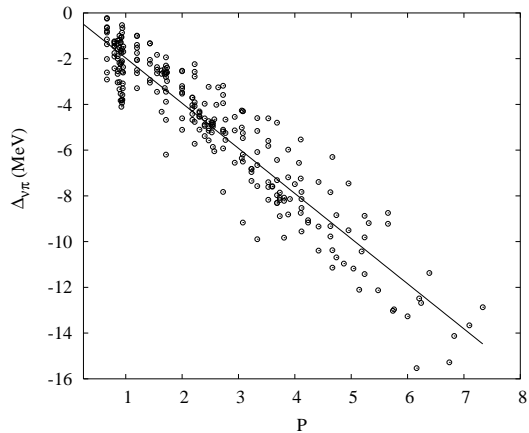}}
\caption{$\Delta_{\nu\pi}$ as a function of $P$ in odd-even nuclei.
\label{bediffgold}}
\end{figure}

\begin{table}[h]
\caption{
The magic proton and neutron numbers used to calculate $N_p$ and $N_n$ for 
nuclei in different mass regions.
\label{list}}
\center
\begin{tabular}{cllc}
 & \multicolumn{1}{c}{$Z$-range} & $N$-range & Core($Z,N$)\\\hline
 & 21 - 23 & 26 - 34 & 20, 28\\
 & 25 - 37 & 30 - 40 & 38, 40\\
 & 35, 37  & 42 - 50 & 38, 50\\
 & 43 - 55 & 50 - 65& 50, 50\\
 &  49 - 63 & 66 - 96 & 50, 82\\
 & 71 - 87 & 104-142 & 82, 126\\
\hline
\end{tabular}
\end{table}
In figure \ref{bediffgold}, we plot the results for a large number of odd-even 
nuclei, lying between mass 47 ($Z=21$) and mass 229 ($Z=87$) as shown in table 
\ref{list}. The results have been plotted only for the nuclei whose experimental
binding energies are available.  The mass regions of table \ref{list} are nearly 
identical with the corresponding ones in Ref. \cite{PLB0}. The experimental 
binding energy 
values in closed neutron shell nuclei between  $Z=49$ and $Z=55$ (with $N=50$
as magic number), and $Z=71$ and $Z=79$ (with $N=126$ as magic number) are not known. 
Following  \cite{PLB0}, we have assumed the $B_{corr}(Z)$ values to be 
identical for $N=50$ and 82 for the nuclei with $Z=49-55$ and $N=51-65$. For 
the nuclei with $Z=71-79$,  we have calculated the $B_{corr}(Z)$ values from
the straight line in Ref. \cite{PLB0} obtained for even-even nuclei 
in this mass region. A total of 278 nuclei have been included in our 
calculation.

Similar to our observation for even $Z$ nuclei, we find that the points lie 
very close to a straight line if plotted as a function of the Casten factor 
Thus, $\Delta_{\nu\pi}$ may be expressed as simply proportional to $P$. One can 
fit a straight line \be \Delta_{\nu\pi}=aP\ee with $a=-1.973\pm 0.024$ MeV
with rms deviation 1.126 MeV. The fitting is for 252 nuclei and does not 
include the values for nuclei with $P=0$ which are defined to be zero.  The 
fitted line has been 
shown in figure \ref{bediffgold}. 

\begin{figure}[t]
\center
\resizebox{7.4cm}{!}{\includegraphics{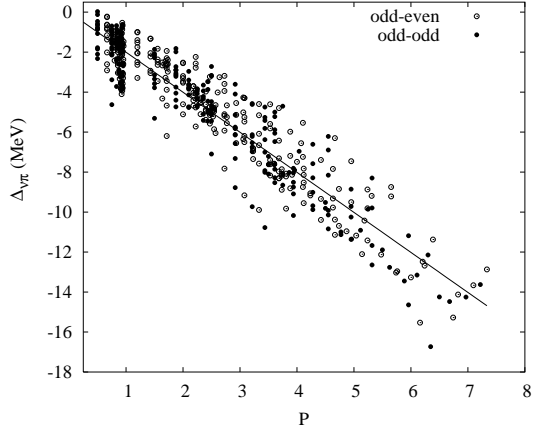}}
\caption{$\Delta_{\nu\pi}$ as a function of $P$ for odd-even and odd-odd 
nuclei. The values for odd-odd nuclei have been shifted as described in the 
text.\label{all}}
\end{figure}

We next turn our attention to odd-odd nuclei. The $B_{corr}(Z)$
values are already known from the study of the odd-even chains. We have 
studied the odd-odd nuclei within the ranges given in table \ref{list}. 
In no case we have  modified the  $B_{corr}(Z)$
values for odd-$N$ isotopes. In our calculation, we ignore the fact that 
the unpaired neutron  actually occupies a particular single particle state,
and breaks the symmetry. However, it is known that the effect of this 
correction to the binding energy is small and is included in the odd-even mass 
difference. 
The results again show a similar trend for odd-odd nuclei. Keeping the 
odd-even mass difference term in the semiempirical mass formula in mind, we 
fit the results using a simple function of the form 
 \be \Delta_{\nu\pi}=aP+\frac{d}{\sqrt{A}}\ee
in terms of the Casten factor and the mass number 
of the nuclide. 
A least square fitting procedure gives the values as 
$a=-2.035\pm0.042$ MeV and $d=10.48\pm1.30$ MeV with a standard deviation of 1.138 MeV
for 245 nuclei. As in the case of even-$Z$ nuclei, we find that the 
coefficients for the Casten factor $P$ for odd-even and odd-odd nuclei are 
identical within errors. The value for $d$ is nearly the same as the 
corresponding coefficient in semi-empirical mass formula, i.e. 11 MeV. In 
figure \ref{all}, the results for all the nuclei described so far, except the 
ones with $P=0$, have been plotted. The results for the odd-odd nuclei have 
been shifted by the amount $-10.48/\sqrt{A}$. A least square fit of the points 
using eqn. (2) leads to a value, $a=-2.003\pm0.017$, with rms deviation of 
1.134 MeV and have also been shown in the form of a straight line. Figure \ref{all} clearly demonstrates once
 again that the n-p interaction is the dominating factor in the correction to 
the RMF binding energy. The odd-even mass difference
may also be expressed  as inversely proportional to mass number, but we
find that in this case, the former prescription fits the data better.
\begin{figure}[b]
\center
\resizebox{7.4cm}{!}{\includegraphics{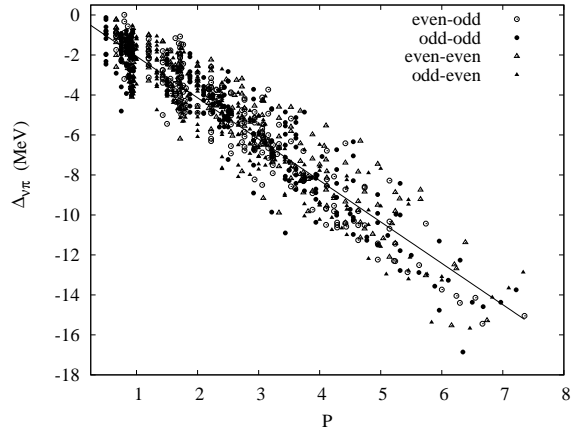}}
\caption{$\Delta_{\nu\pi}$ as a function of $P$. See text for details.\label{allis}}
\end{figure}

The results for even-$Z$ nuclei from Ref. \cite{PLB0} and for odd-$Z$ nuclei in
the present work  are clearly seen to follow the same linear pattern. In order 
to express all the values with a single relation, we have assumed the form  
in eqn (3). Least square fitting yields the values
for the parameters as $a=-2.070\pm0.015$ MeV and, in odd-$N$ nuclei, 
$d=12.05\pm 0.70$ MeV
with an average deviation of 1.132 MeV for 932 nuclei.
The constant $d$ is taken to be zero in even-neutron isotopes. 
The results for nuclei with $P\neq 0$, after shifting the values for odd neutron nuclei taking the second 
term in eqn (3) into account, have been plotted in figure \ref{allis} where
the straight line represents the equation $\Delta_{\pi\nu}=-2.070P$.

The corrections derived in the present procedure  may be employed to improve 
the agreement 
between the calculated and experimental binding energy values. The present mean 
field calculation is a spherical one, and does not take deformation  
into account. It is expected to underpredict the binding energy far away
from the closed shell. However, with the corrections, it is 
possible to obtain an agreement comparable to or even better than the values 
calculated using a deformed mean field approach. Thus the present approach may 
be very useful in predicting the mass of nuclei far from the stability valley.

\begin{figure}[h]
\center
\resizebox{7.4cm}{!}{ \includegraphics{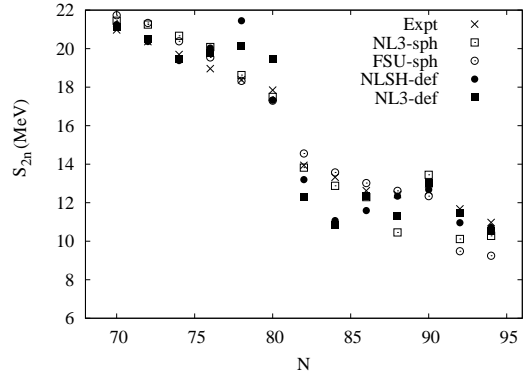}}
\caption{Two neutron separation energy in Nd in various approaches.
See text for details.\label{2nnd}}
\end{figure}

As an example of comparison with RMF calculations which take
deformation explicitly into account, we have chosen the Nd 
isotope chain and studied the two neutron separation energy values for even-even
isotopes. The results are plotted in figure \ref{2nnd}. Values obtained from both 
the FSU Gold and the NL3 spherical calculations have been corrected using the 
method described in the present work and Ref. \cite{PLB0}. The deformed NLSH results are from 
Lalazissis {\em et al} \cite{lalazissis}. The deformed NL3 calculations use a 
deformed harmonic oscillator basis using gap parameters obtained from odd-even 
mass difference whenever available. In $^{130,156}$Nd, the values have been 
calculated using the empirical prescription $11.2/\sqrt{N(Z)}$. We find that the
agreement using the present approach is comparable to or sometimes better than 
that observed in the deformed calculations.

One obvious shortcoming in the present procedure is that the odd-even mass
difference has been taken care of by a global relation while the actual 
quantity may be more appropriately described in a proper mean field procedure. 
It may be more appropriate to consider the effect of odd-even mass difference 
in a fully microscopic approach.

\section{Summary}

To summarize, the differences between the experimental and the theoretically 
calculated binding energies in RMF approach have been 
calculated for a large number of odd-$Z$ nuclei from $A=47$ to 229. 
The n-p interaction is expected to be the major contributor to the difference 
between the theoretical and the experimental binding energies in RMF. 
This difference, excluding certain mass regions and taking different effects
into account, may be linearly parametrized by the 
Casten factor, a commonly used measure of n-p interaction.
The trend is similar to the case of even-$Z$ nuclei.

\section*{Acknowledgment}

This work is carried out with financial assistance of the UGC sponsored
DRS Programme of the Department of Physics of the University of Calcutta.

\newpage

\end{document}